# Structural Modulation in LaO$_{0.9}$F$_{0.1}$BiSe$_2$ Single Crystals Revealed by Scanning Tunneling Microscopy/Spectroscopy

**N Ishida, S Demura, Y Fujisawa, S Ohta, K Miyata and H Sakata**

Department of Physics, Tokyo University of Science, Shinjyuku-ku, Tokyo 162-8601, Japan

E-mail: 1215605@ed.tus.ac.jp

**Abstract**. We present scanning tunneling microscopy and spectroscopy measurements on a cleaved surface of the LaO$_{0.9}$F$_{0.1}$BiSe$_2$ single crystals. Tunneling spectra show a finite local density of states at $E_F$, which is consistent with metallic conductivity in bulk. In addition, the existence of the supermodulation running along the diagonal directions of Bi square lattice was revealed. The period of the supermodulation was about 3 to 5 times the length of the lattice constant. This period is close to that observed in LaO$_{0.5}$F$_{0.5}$BiSe$_2$.

## 1. Introduction

Enormous attentions have been paid to layered superconductors, such as cuprates and iron-based superconductors because of their high superconducting transition temperature $T_c$. Newly discovered BiS$_2$-based superconductors $Ln$O$_{1-x}$F$_x$Bi$Ch_2$ ($Ln$ = La, Pr, Ce, Nd, Yb and Bi, $Ch$ = S, and Se) also have a layered structure [1]. A lot of BiS$_2$-based superconductors were synthesized with several kinds of $Ln$ or $Ch$, where the carrier density is controlled by the amount of F.

Recently, scanning tunneling microscopy and spectroscopy (STM/STS) measurements on single crystals of NdO$_{0.7}$F$_{0.3}$BiS$_2$ revealed the existence of a novel electronic structure so-called "Checkerboard stripe" structure in BiS plane. However, spectroscopic gap was observed at the Fermi energy $E_F$ despite of metallic bulk properties [2]. Furthermore, the unexpected supermodulation with the period of about 5 times the length of the lattice constant running along the diagonal direction of the Bi square lattice was revealed in LaO$_{0.5}$F$_{0.5}$BiS$_2$ ($x$ = 0.5) single crystals by synchrotron X-ray measurements [3] and in LaO$_{0.5}$F$_{0.5}$BiSe$_2$ ($x$ = 0.5) single crystals by STM/STS measurements [4]. Theoretically, in high F doping samples, such as LaO$_{0.5}$F$_{0.5}$BiS$_2$ ($x$ = 0.5), the phonon instability spreading from near Γ point to M point is predicted while in low F doping samples, such as LaOBiS$_2$ ($x$ = 0), the phonon instability around Γ point is predicted [5]. Thus, the measurements in different F doping samples give us the cue to understand the nature of the observed supermodulation. In this paper, we performed STM/STS measurements in LaO$_{0.9}$F$_{0.1}$BiSe$_2$ ($x$ = 0.1) single crystals and observed the supermodulation. The results were compared with the supermoduation reported in LaO$_{0.5}$F$_{0.5}$BiSe$_2$ ($x$ = 0.5).

## 2. Experimental

Single crystals of $LaO_{0.9}F_{0.1}BiSe_2$ were grown by CsCl flux method in vacuumed quartz tubes [6]. Mixtures of Bi (Mitsuwa Chemicals Co. Ltd., 99.9%), Se (Kojyundo Chemicals Co. Ltd., 99.99%), $Bi_2O_3$ (Kojyundo Chemicals Co. Ltd., 99.99%), and $BiF_3$ (Stella Chemifa Co. Ltd., 99.9%) were ground with nominal compositions of $LaO_{0.9}F_{0.1}BiSe_2$ except La. After grind into powder, La ribbons (Rare Metallic Co. Ltd., 99.9%) were mixed into the mixture. The mixture of 0.8 g was mixed with CsCl powder of 5 g, and sealed in an evacuated quartz tube. The thermal process is the followings. First, the tube was heated up to 800 °C at a rates of 100 °C/h, secondly, kept for 10 hours, finally, cooled down to 630 °C at a rates of 10 °C/h. After this thermal process, the heated material was washed by distilled water to remove the flux.

Superconducting transition temperatures $T_c$ were determined from the temperature dependence of the resistivity. $T_c$ of the sample of $x = 0.1$ and $x = 0.5$ were 3.2 K and 3.8 K, respectively. The difference in $T_c$ indicates the difference in the actual amount of F between $x = 0.1$ and $x = 0.5$ samples [1].

STM/STS measurements were performed with laboratory-built STM in the He gas at 4.2 K. The clean surface of $LaO_{0.9}F_{0.1}BiSe_2$ single crystal was prepared by cleavage at 4.2 K *in situ*. A bias voltage was applied to the sample in all measurements. Tunneling spectra were obtained by numerically differentiation of the *I-V* curves.

## 3. Result

Figure 1 (a - c) show typical STM images and a tunneling spectrum in $LaO_{0.9}F_{0.1}BiSe_2$ ($x = 0.1$) single crystal. The STM image in Figure 1 (a), which was taken at a positive sample bias voltage, indicates a square lattice with a period of $a_0 = 4.7$ Å corresponding to the in-plane lattice constant. The observed bright and dark spots are Bi atoms and defects of Bi atoms, respectively [2, 4, and 7]. In addition, some Bi defects have dark streaks along one or two diagonal directions of the Bi square lattice. The density of defects was about 1% and was almost the same as that in $x = 0.5$ sample [4]. Thus, the Bi defect density is almost independent of the amount of F.

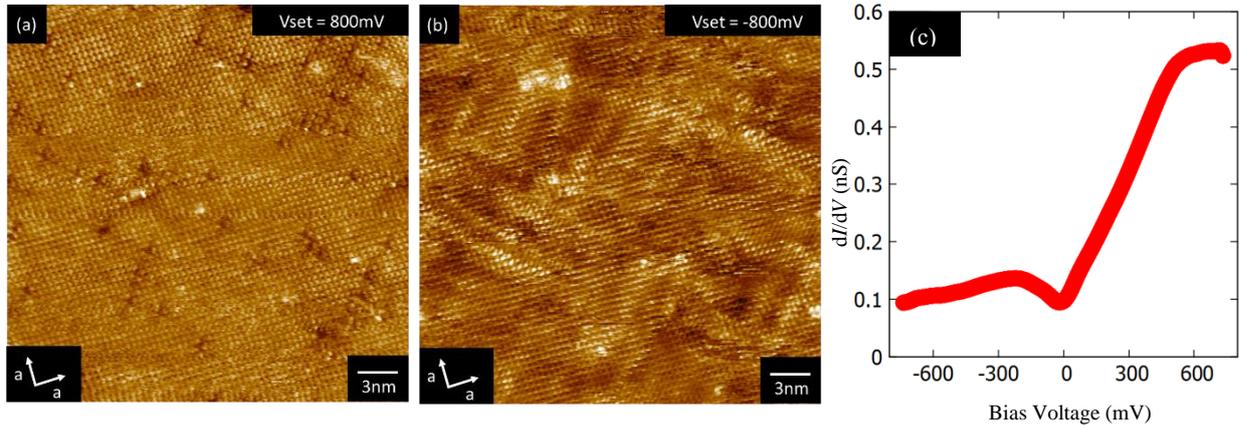

Figure 1 (Color online). (a) STM image of $LaO_{0.9}F_{0.1}BiSe_2$ ($x = 0.1$) single crystals taken at $V_{set} = 800$ mV and $I_{set} = 300$ pA. (b) STM image taken at $V_{set} = -800$ mV and $I_{set} = 300$ pA. (c) Typical tunneling spectrum taken at $V_{set} = 800$ mV and $I_{set} = 300$ pA.

Figure 1 (b) shows the STM image taken at a negative sample bias voltage for the same field of view as Figure 1 (a). In addition to the Bi square lattice, the supermodulation running along the diagonal directions of the Bi square lattice exists. Figure 2 (a) shows the low-pass-filtered image of Figure 1 (b) superimposed with the positions of Bi defects and dark streaks seen in Figure 1 (a). Bi defects and streaks are marked by gray circles and red dashed lines, respectively. The observed supermodulation is not two dimensional wave spreading over the surface uniformly, but consists of locally developed one dimensional waves. There seems to exist the relation between the supermodulation and Bi defects. There are two tendencies: (i) the supermodulation which is perpendicular to a dark streak terminates at Bi defects, (ii) dark streaks are located in the boundaries where the direction of supermodulation changes. This indicates that Bi defects affect the modulation seriously.

Next, we focus on the period of the supermodulation. Because the observed modulation is not regular, Fourier transformed images, shown in Figure 2 (b) and 2 (c), indicate that the signal corresponding to the supermodulation is no longer spots but broadened. Thus, the period was estimated from the real space image shown in Figure 1 (b).

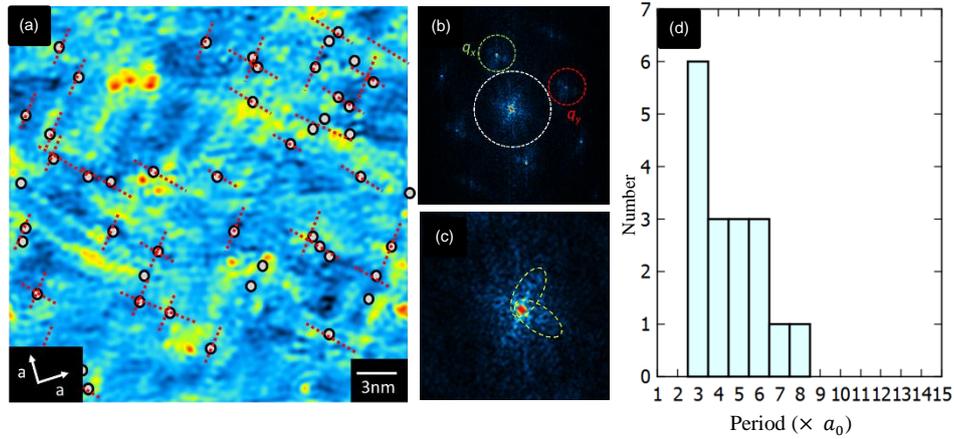

Figure 2 (Color online). (a) Low-pass-filtered image of Figure 1 (b). Cut off wave number is $|q| = 0.7 \times 2\pi/a_0$ marked by white dashed circle in Figure 2 (b). (b) Fourier transform image of Figure 1 (b). (c) Magnified image of Figure 2 (b) around $\Gamma$ point. (d) Histogram of the period of the modulation.

To evaluate the period of the supermodulation, we took cross sections of the STM image and measured the distance between the crests of the modulation at several locations. Here, we call this distance as period. The histogram of the period is shown in Figure 2 (d). According to the Figure 2 (d), the estimated period was distributed between 1 and 2 nm, which corresponds to 3 to 5 times the length of the lattice constant. The period of 3 times the length of the lattice constant is the most frequent. Thus the observed period shows a little shorter value than that in $LaO_{0.5}F_{0.5}BiSe_2$ ($x = 0.5$).

## 4. Discussion

The STM observations revealed the existence of the supermodulation even in $LaO_{0.9}F_{0.1}BiSe_2$ ($x = 0.1$). Likewise $LaO_{0.5}F_{0.5}BiSe_2$ ($x = 0.5$), the observed supermodulation runs along the diagonal directions of Bi square lattice. Therefore, it is considered that the supermodulation in $LaO_{0.9}F_{0.1}BiSe_2$ ($x = 0.1$) is essentially the same as that in $LaO_{0.5}F_{0.5}BiSe_2$ ($x = 0.5$). Although the period of the observed supermodulation in $LaO_{0.9}F_{0.1}BiSe_2$ was a little shorter than that observed in $LaO_{0.5}F_{0.5}BiSe_2$ ($x = 0.5$), their features are almost same. Thus, it is difficult to say that the modulation is significantly modified with the amount of F. In theoretical calculation, doping dependence of the wave vector at which the phonon becomes instable in $LaO_{1-x}F_xBiS_2$ is predicted [5]. The observed results, however, does not consistent with the theoretical expectation. Thus, the origin of the formation of the supermodulation is still an open question.

It is noted, as shown in Figure 1 (c), the observed tunneling spectrum shows a finite density of states at the Fermi energy. This is consistent with the electric resistivity measurements [6, 8], indicating that the STM/STS measurements in this material represent the bulk property of the sample. Thus, the observed supermodulation is thought to also exist in the bulk. In fact, recent reports on synchrotron X-ray diffraction measurements in $LaO_{0.5}F_{0.5}BiS_2$ [3] claimed the existence of such supermodulation.

## 5. Summary

We succeeded in STM/STS observation in $LaO_{0.9}F_{0.1}BiSe_2$ single crystals and revealed the existence of the supermodulation running along the diagonal direction of the Bi square lattice at a negative bias voltage. Although there is fluctuation in the period of the supermodulation, the period was about from 3 to 5 times the length of the lattice constant, which is close to the value observed in $LaO_{0.5}F_{0.5}BiSe_2$ single crystals.